\documentclass{aa}  

\usepackage{graphicx}
\usepackage{txfonts}
\usepackage{comment}
\usepackage{natbib}
\usepackage{verbatim}
\usepackage{fancyvrb}

\renewcommand{\vec}[1]{\mathbf{#1}}                                                                                     
\newcommand{\earth}{\text{earth}}                                                                                                       
\renewcommand{\sun}{\text{sun}}                                                                                                         
\newcommand*{\norm}[1]{\left\lVert#1\right\rVert}                               
\newcommand*{\abs}[1]{\left\lvert#1\right\rvert}                                        

\usepackage{ulem}

\usepackage{color}

\begin{document}

   \title{LEADER: fast estimates of asteroid shape elongation
and spin latitude distributions from scarce photometry}
        \titlerunning{LEADER}


   \author{H. Nortunen\inst{1} and M. Kaasalainen\inst{1}}

   \institute{Tampere University of Technology, Department of Mathematics, 
PO Box 553, 33101 Tampere, Finland              
             }

   \date{Received 13 June 2017; accepted 27 September 2017}

\authorrunning{H. Nortunen \& M. Kaasalainen}

 
  \abstract
   {Many asteroid databases with lightcurve brightness measurements (e.g. WISE, Pan-STARRS1) contain enormous amounts of data for asteroid shape and spin modelling. While lightcurve inversion is not plausible for individual targets with scarce data, it is possible for large populations with thousands of asteroids, where the distributions of the shape and spin characteristics of the populations are obtainable.}
   {We aim to introduce a software implementation of a method that computes the joint shape elongation $p$ and spin latitude $\beta$ distributions for a population, with the brightness observations given in an asteroid database. Other main goals are to include a method for performing validity checks of the algorithm, and a tool for a statistical comparison of populations.}
   {The LEADER software package read the brightness measurement data for a user-defined subpopulation from a given database. The observations were used to compute estimates of the brightness variations of the population members. A cumulative distribution function (CDF) was constructed of these estimates. A superposition of known analytical basis functions yielded this CDF as a function of the (shape, spin) distribution. The joint distribution can be reconstructed by solving a linear constrained inverse problem. To test the validity of the method, the algorithm can be run with synthetic asteroid models, where the shape and spin characteristics are known, and by using the geometries taken from the examined database.}
   {LEADER is a fast and robust software package for solving shape and spin distributions for large populations. There are major differences in the quality and coverage of measurements depending on the database used, so synthetic simulations are always necessary before a database can be reliably used. We show examples of differences in the results when switching to another database.}
{}


   \maketitle
%

\section{Introduction}

Photometric observations of total (disk-integrated) brightnesses are by far the most abundant source of information on asteroids in the sense of population coverage (\v{D}urech et al. 2015). There are currently some thousand shape and spin models from photometry for individual asteroids -- for example, see the the Database of Asteroid Models from Inversion Techniques (DAMIT\footnote{http://astro.troja.mff.cuni.cz/projects/asteroids3D/web.php.}) site -- and tens of thousands more will be obtained from sparse photometry from various sky surveys such as Panoramic Survey Telescope and Rapid Response System (Pan-STARRS), Gaia, and Large Synoptic Survey Telescope (LSST). Infrared photometry from, for example, the Wide-field Infrared Survey Explorer (WISE) mission can also be added to visual data to obtain size and surface material parameters for thousands of targets (\v{D}urech et al. 2016). 

Even larger population-level attributes can be obtained by using all available photometric measurements from the rich survey databases also when the data are not sufficient for individual models, as shown in Nortunen et al. (2017) and Cibulkov{\'a} et al. (2017). We refer to such data as scarce photometry to distinguish it from sparse photometry. The possibility of obtaining population-level attributes is based on the principle of using even a few data points per target to construct a set of brightness variation estimates for a given population (defined by the user). As shown in Nortunen et al. (2017), such estimates are sufficient for robustly reconstructing the shape elongation and (ecliptically symmetric) spin latitude distributions of the population when there are thousands of samples available. The determination of the spin distribution requires the population to be dominated by orbits concentrated near the ecliptic plane.

In this paper, we describe in detail the software package Latitudes and Elongations of Asteroid Distributions Estimated Rapidly (LEADER) for obtaining the shape and spin distributions. The mathematical principles are presented in Nortunen et al. (2017), but the computational aspects require further exposition given here. The main issues are the choice of the grid discretization level in solving the inverse problem, experimenting with various setup choices to establish the stability of the result, the `deconvolution' of the result distributions in the shape-spin plot plane, and the mandatory use of simulations (from synthetic data for the same observing geometries as in the input database) to check the reliability of the inversion. We emphasize that, while the method itself is quite robust in the sense that the inaccuracy of the underlying ellipsoidal model is well tolerated since only the large-scale elongation and spin estimates are obtained, the properties and coverage of the database entirely dictate the reliability of the results.

This paper is organized in the following manner. In Sect. 2 we present the main algorithm, some computational details, and the visualization by deconvolution, while in Sect. 3 we discuss the implementation of the necessary simulations from synthetic data and the comparison routines between populations. In Sect. 4 we illustrate the use of the LEADER package with examples. We sum up in Sect. 5, and in the Appendix we describe some main components of the software.

\section{Main algorithm for computing distributions} \label{sec:leader-main}

The main algorithm consisted of roughly three phases. First, we had a forward model for computing the brightness variation estimate $\eta$ from observations. Then, we considered the inverse problem and determined the joint $(p, \beta)$ distribution. Finally, we plotted the results and applied a deconvolution filter to smoothen the solution.

\subsection{Forward model: brightness variations} \label{sec:forward}

We considered a population of $N$ asteroids. For our analysis, we required a large population, with $N \gtrsim 1000$. In the forward model, our observable is the brightness deviation estimate $\eta$. Our objective is to obtain one or more $\eta$s for each asteroid. With the observed brightness values $L$, we defined $\eta$ as
        \begin{equation}
        \eta = \frac{\Delta (L^2)}{\langle L^2\rangle} ,
        \label{eq:eta-def}
        \end{equation}
        where $\Delta(L^2)$ is a measure of variation for $L^2$ as defined in \citet{hn-method}:
        \[ \Delta (L^2) = \sqrt{\langle(L^2-\langle L^2\rangle)^2\rangle} . \]
We used the squared brightness $L^2$ for convenience; a more detailed explanation is given in \citet{hn-method}. We were then able to compute the amplitude $A$ from each $\eta$, and from all the amplitude values of the population, we constructed the cumulative distribution function (CDF) $C(A)$. To obtain the brightness deviation estimates, we analyzed the data file of asteroid $i$, where $i=1$, $\ldots$, $N$, repeating the following steps for each file:
\begin{enumerate}       
        \item From the data file, we read the Julian date, intensity, and the directions of the Earth and Sun (these should be computed if not given) as seen from the asteroid translated into the origin. We denoted the direction vectors of the Earth and Sun $\vec{e}_{\earth}$ and $\vec{e}_{\sun}$, respectively. For each data point, we computed the phase angle $\alpha$ between the Sun and the Earth, and required that
        \begin{equation}
        \alpha := \arccos( \vec{e}_{\sun} \cdot \vec{e}_{\earth} ) \le \alpha_{\mathrm{tol}} .
        \label{eq:angtol}
        \end{equation}
        In Nortunen at al. (2017), we used a limit of $\alpha_{\mathrm{tol}} = 30^{\circ}$ for the WISE database. For the Pan-STARRS1 database, we used $\alpha_{\mathrm{tol}} = 20^{\circ}$. We filtered out each brightness measurement where the phase angle exceeds this given tolerance. In addition, we required that a set of measurements has been done within a small enough change in geometry, and at least five brightness values are required for a valid $\eta$ estimate. For Pan-STARRS1, the phase angle $\alpha$ changes rapidly in time, so we could have used a condition that all measurements in a set are done within, for example, three days, to keep the change in the phase angle small. For the WISE database, more liberal rules can be used with the time span, as the phase angle changes at most $0.4^{\circ}$ within a one-week-long set of measurements. With this set of rules, we separated the measurements on a single data file into multiple sets. After that, we analyzed each set individually for computing $\eta$.
        
        \item \textit{Phase angle correction.} Depending on the phase angle $\alpha$, we should apply either an exponential or a linear correction to the brightness values $L(\alpha)$, as explained in \citet{icarus}. Let us consider a single set of measurements. If
        $\min_i \alpha_i < 8^{\circ}$
        in the set, we required that the phase angle changes no more than one degree, that is:
        \begin{equation} \label{eq:phasechange01} (\max_i \alpha_i) - (\min_i \alpha_i) \le 1^{\circ}. \end{equation}
        If the phase angle does change for more than one degree, we applied an exponential correction. In other words, we applied an exponential curve $a \exp(b\alpha)$ into the $(\alpha, L)$ data, and used it to normalize the brightness data into the form
        \begin{equation} \label{eq:expcorr} \frac{L(\alpha)}{a \exp(b\alpha)} , \quad a > 0, \ b < 0 . \end{equation}
        If $\min_i \alpha_i \ge 8^{\circ}$ in the set, we required that the phase angle changes at most two degrees:
        \begin{equation} \label{eq:phasechange02} (\max_i \alpha_i) - (\min_i \alpha_i) \le 2^{\circ}. \end{equation}
        If the phase angle changes more, we applied a linear correction by using a least squares fit of the form $(a\alpha + b)$ into the $(\alpha, L)$ data, and use it to normalize the brightness data into
        \begin{equation} \label{eq:lincorr} \frac{L(\alpha)}{a\alpha + b} , \quad a < 0 . \end{equation}
        We note that with a small number of points, the correction may be unstable and can be omitted.
        
        \item For each measurement set, we computed the brightness variation $\eta$ using Eq.\ \eqref{eq:eta-def}. In \citet{hn-method}, we derived how $\eta$ and the amplitude $A$ are directly related:
        \begin{equation}
        A=\sqrt{1-\Big(\frac{1}{\sqrt{8}\eta}+\frac{1}{2}\Big)^{-1}} .
        \label{eq:eta-A}
        \end{equation}
  We omitted any complex-valued or non-finite amplitudes.
\end{enumerate}

When all the brightness variations and amplitudes ($\eta$ and $A$) have been computed, we sorted the amplitudes in an increasing order. Then the CDF of $A$ is simply $C(A_i) = i/k$ for $i=1$, $\ldots$, $k$.

\subsection{Inverse problem: obtaining the joint distribution for latitudes and elongations} \label{sec:inverse}

In the inverse problem, our objective is to determine the distributions of two parameters, the shape elongation $p \in [0, 1]$ and the spin latitude $\beta \in [0, \pi/2]$. We modelled the asteroids with the shape of a triaxial ellipsoid, with semiaxes $a$, $b$ and $c$, and $a \ge b = c = 1$. With this model, the shape elongation is simply $p = b/a$. Here a small $p$ value corresponds to an elongated body, while $p = 1$ describes a sphere. For the spin latitude, $\beta = 0$ represents a spin direction that is perpendicular to the ecliptic plane, while $\beta = \pi/2$ means the spin direction is in the ecliptic plane\footnote{We note that in our convention, $\beta$ is the complementary angle of the traditionally used $\beta$. This is due to mathematical purposes. To avoid confusion, we use radians instead of degrees for the values of our $\beta$ in plots.}. Our model does not have any way of distinguishing whether the spin latitude is above or below the ecliptic plane. We assumed that the observations are concentrated near the ecliptic plane. This assumption does not usually hold entirely; the amount of variation in the ecliptic latitudes of the observations depends on the orbits of the population and the sampling epochs in database used. As a result, the computed $\beta$ distribution tends to be less accurate than the $p$ distribution. We tested the validity of the assumption, and for the databases we used in this paper, some $95\%$ of the observations were concentrated within a $\pm 20^{\circ}$ sector near the ecliptic plane.

To solve the inverse problem, we created a grid of $N_p \times N_{\beta}$ bins for our $(p, \beta)$ values. For our algorithm, we typically used $N_p = 20$ and $N_{\beta} = 29$, so every bin is approximately $0.05 \times 0.05$ units in size, and equally spaced. We chose $(p_i, \beta_j)$ as a random point near the centre of each bin. Alternatively, as the values $p < 0.4$ are expected to have lower occupation numbers than the higher $p$ values, we may lower the resolution for such values. Similarly, since the occupation numbers for $\beta$ are expected to be somewhat proportional to $\sin \beta$ (this means a uniform density on a sphere), low $\beta$ values were expected to have lower occupation numbers, so we may lower the resolution of, for example, $\beta$ values smaller than $\pi/4$.

In \citet{hn-method}, we derived how the CDF $C(A)$ can be expressed in an analytical integral form when we have infinite observations available in every geometry. When we used a grid of points $(p_i, \beta_j)$, we were able to express $C(A)$ as the following superposition:
\begin{equation}
C(A) = \sum_{ij} w_{ij} \, F_{ij}(A) .
\label{eq:lineq00}
\end{equation}
Here $w_{ij}$ are the occupation numbers (weights) of each bin $(p_i, \beta_j)$, and
        \begin{equation}
F_{ij}(A)=\left\{\begin{array}{rl}
0, & A\le p_i\\
\frac{\pi}{2}-\arccos\frac{\sqrt{A^2-p_i^2}}{\sin\beta_j\sqrt{1-p_i^2}} ,& p_i<A<\mathcal{F}(p_i, \beta_j)\\
\frac{\pi}{2}, & A\ge\mathcal{F}(p_i, \beta_j)
\end{array}\right.\label{Fijeq}
\end{equation}
are analytical basis functions, where $\mathcal{F}(p_i, \beta_j) = \sqrt{\sin^2\beta_j+p_i^2\cos^2\beta_j}$.

Next, we constructed the data matrix $M$ (with $k$ rows and $N_p \cdot N_{\beta}$ columns) such that each column of $M$ contains a basis function $F_{ij}(A)$. When we write $C(A) =: C \in \mathbb{R}^k$, the superposition of Eq.\ \eqref{eq:lineq00} can be written as a linear system,
\begin{equation}
Mw=C,
\label{eq:lineq}
\end{equation}
where the unknown vector $w \in \mathbb{R}^{N_p \cdot N_{\beta}}$ contains the occupation numbers $w_{ij}$ of each bin. Before solving the system, the use of regularization is highly recommended, especially for the more unstable $\beta$. Let $R_p$ be an $\big((N_p-1) \cdot N_{\beta}\big) \times (N_p \cdot N_{\beta})$ matrix that is meant to smooth the solution for $p$, and $R_{\beta}$ be the respective $\big(N_p \cdot (N_{\beta}-1)\big) \times (N_p \cdot N_{\beta})$ regularization matrix for $\beta$. For indices $ij$, we have:
$$
(R_p)_{ij} = \left\{ \begin{array}{rl}
-1/(p_{i+1}-p_i), & i=j \\
1/(p_{i+1} -p_i), & j=i+1 \\
0, & {\rm elsewhere},
\end{array} \right.
$$
and similarly for $R_{\beta}$. The regularization matrices approximate the gradients at each $w_{ij}$ in the $p$- and $\beta$-directions. For the regularization parameters, we typically used values $\delta_p = 0.1$ and $\delta_{\beta} = 1$. Now, let us create an extended matrix $\tilde M$ and an extended vector $\tilde C$:
$$
\tilde M=\left(\begin{array}{r}
M\\
\sqrt\delta_p R_p\\
\sqrt\delta_{\beta} R_{\beta} \end{array}\right), \quad \tilde C=\left(\begin{array}{l}C\\0_{(N_p-1)N_{\beta}}\\
0_{N_p(N_{\beta}-1)}\end{array}\right),
$$
with our extended linear system being
\begin{equation}
\tilde M w = \tilde C .
\label{eq:lineq-ext}
\end{equation}
To obtain the occupation numbers $w_{ij}$, we solved for $w$ from Eq.\ \eqref{eq:lineq-ext} by using, for example, MATLAB's \citep{matlab} linear least squares method with a positivity constraint $w_{ij} \ge 0$. The peak of the joint $(p, \beta)$ distribution is simply the $(p_i, \beta_j)$ bin with the highest occupation number $w_{ij}$.

\subsection{Visualization} \label{sec:plots}

To estimate the goodness of the fit $C(A) = \sum_{ij} w_{ij} F_{ij}(A)$ from Eq.\ \eqref{eq:lineq00}, or the equivalent form $C = Mw$ from Eq.\ \eqref{eq:lineq}, we may plot $C$ and $Mw$ in the same plot. The relative error $\norm{C-Mw}/\norm{C}$ is usually less than $1\%$ when the population contains at least 1000--2000 objects. For the actual joint distribution $f(p, \beta)$, where $f(p_i, \beta_j) = \tilde w_{ij}$ and $\tilde w_{ij}$ are the occupation numbers $w_{ij}$ normalized such that $\sum_{i, j} f(p_i, \beta_j) = 1$, we may plot the solution $(p, \beta, f(p, \beta))$ either as a three-dimensional surface plot, or alternatively as a contour plot. The marginal density functions (DFs) can also be computed for $p$ and $\beta$:
\begin{equation}
f(p_i) = \sum_{j=1}^{N_{\beta}} \tilde w_{\rm ij} , \quad f(\beta_j) = \sum_{i=1}^{N_p} \tilde w_{\rm ij} .
\label{eq:marginal}
\end{equation}

As a post-processing tool, we may apply deconvolution to correct for dispersion in the obtained solution. The deconvolution is used as a primarily visual tool, and it is applied only for the joint $(p, \beta)$ distribution, not the marginal DFs. In order to know what kind of post-solution correction is needed, synthetic simulations (Sect. \ref{sec:synth}) should be performed on the database used. With the synthetic simulations, we gained understanding of the accuracy levels of the method, and were able to detect systematic errors associated with the database. As the solution tends to spread when moving away from the peak, it is a common procedure to introduce dampening to bins away from the peak. Let $i^*$ and $j^*$ be the indices for the statistical peak of the solution, that is, the bin with the highest occupation number. Then, the dampening we applied is
\begin{equation}
\tilde w_{ij}^{\rm corr} = \frac{\tilde w_{ij}}{ (\abs{i^* - i} + \abs{j^* - j} + 1)^n } .
\label{eq:deconvo-damp}
\end{equation}
Usually it suffices to choose $n=1$, but if heavier dampening is required, we may choose a larger $n$. As the solution of the shape elongation $p$ is often shifted too much to the left (towards more elongated values), for example, by the amount $\Delta \mathcal{P} \ge 0$, we may additionally shift the $p$ values to the right (towards more spherical values):
\begin{equation}
p_i^{\rm corr} = \min(p_i + \Delta \mathcal{P}, 1) .
\label{eq:deconvo-p}
\end{equation}
For the WISE database, we chose $\Delta \mathcal{P} = 0.1$ in \citet{hn-method}. The more noisy the database is, the higher the required shift $\Delta \mathcal{P}$ is. For $\beta$, the error behaviour is much harder to model and may lead to exaggerated correction, so typically we did not apply any correction in the $\beta$ direction. However, we acknowledge that the solution tends to avoid extreme ends, so values near $\beta = 0$ (perpendicular to the ecliptic plane) and values near $\beta = \pi/2$ (in the ecliptic plane) have a tendency to shift away from the end points, moving towards the middle.

The computation times of the main algorithm are negligible. Depending on the size of the inspected population, reading the geometries from a database may take a few minutes, while the computation of the solution via the inverse problem is even faster. Therefore, it is easy to experiment with different grids for the solution of the inverse problem, or to test different populations. The latter means it is also fast to compare populations, which we will discuss in more detail in Section \ref{sec:comparison}.

\section{Main implementations of LEADER}

In Sect. \ref{sec:synth}, we discuss an essential test when using the LEADER package: accuracy estimation by running simulations on synthetic data. Simulations are the only way to gain an understanding of how applicable the main algorithm is for a given database, and they should be performed on every database before those can be reliably used. In Sect. \ref{sec:comparison} we describe an extension of the main algorithm, an application for comparing shape elongation and spin latitude distributions of two populations.

\subsection{Verifying the method using simulations based on synthetic data} \label{sec:synth}

The only way to test the correctness of the obtained solution is to run simulations based on synthetic data, where the $(p, \beta)$ distribution of the artificial population is known, and see how accurately the solution is obtained. The test can additionally be used to detect systematic errors. The level of accuracy has a strong dependence on the database used \citep{hn-method}, so whenever we start to use a new database, it is necessary to run simulations to see how well our method performs with the database.

The synthetic simulations begin by choosing a single peak for the $(p, \beta)$ distribution. Let us denote this peak $(p^*, \beta^*)$. Now, let us assume we run the simulation for $N$ asteroids in a population. Then, we repeat the following steps for each asteroid $i$, where $i=1$, $\ldots$, $N$:
\begin{enumerate}
        \item We choose an asteroid model from DAMIT\footnote{We have chosen DAMIT as the source for synthetic models doe to its rich variety of realistic shapes.}, with a shape elongation $p=b/a$ (here $a$ is the longest diameter in the equatorial
$xy$-plane, and $b$ is the width in the corresponding orthogonal direction) that is close to the peak value $p^*$. For example, we have set a criterion,
        \begin{equation} \label{eq:p-criterion} \abs{p-p^*} \le 0.075 . \end{equation}
        We may apply basic transformations, such as stretching, on DAMIT objects in order to get the intended shape elongation value for the asteroid. We compute the normal and area for each facet of the body.
        \item The next step is to construct the brightness data and our $\eta$ estimate. We choose a $\beta$ from Gaussian distribution, with $\beta^*$ as the mean value, and $0.05$ as the standard deviation (with a restriction that $\beta \in (0, \pi/2)$). We also fix the longitude $\lambda$ by choosing it from a uniform distribution, $\lambda \in [0, 2\pi]$.
        \item Next, we need geometries from the asteroid database we are studying. We read the data from a data file belonging to that database.  We extract the direction vectors of the Sun and the Earth from the data file, and filter out the cases when the condition of Eq.\ \eqref{eq:angtol} is violated. Next, we transform these vectors to the asteroid-fixed frame using a coordinate transformation \citep{icarus}, with our fixed $\beta$ and $\lambda$. We denote the direction vectors in the asteroid's own frame $\vec{e}_{\sun}$ and $\vec{e}_{\earth}$.
        Then we compute the total brightness $L$ for the target.
        Finally, we add a minor Gaussian perturbation to $L$ to simulate noise.
        \item From now on, we proceed as in Sect. \ref{sec:forward}. We require at least five $L$ values for a valid $\eta$ estimate, and if necessary, we only consider the measurements that have been done within a short time span to keep the changes in geometry small. If necessary, we apply phase correction on the sets, and finally, we compute $\eta$ and $A$ for each set as in Eq.\ \eqref{eq:eta-def} and \eqref{eq:eta-A}. We sort the amplitudes in an increasing order, and construct the CDF $C(A)$ as in Section \ref{sec:forward}. As the actual $p$ and $\beta$ distributions are known with the synthetic data, it is often illustrative to plot the marginal DFs and the contour plot of the joint $(p, \beta)$ distribution.
\end{enumerate}
The next phase is to obtain the solution of the inverse problem. This is done in the same way as that described in Sect. \ref{sec:inverse}. The graphical presentation of the results, as well as possibly applying a deconvolution filter, is done identically to Sect. \ref{sec:plots}.

\subsection{A comparison of two populations} \label{sec:comparison}

A comparison between user-determined populations is typically desirable. Taking into account that the database used tends to cause biases in the `absolute' values of distributions, the relative differences between distributions can be expected to be more robust results from the database.

For populations $S_1$ and $S_2$, we ran the algorithm described in Sect. \ref{sec:leader-main}. We collected the $p$, $f(p)$, $\beta$ and $f(\beta)$ information from both algorithms, where $p$ and $\beta$ are the grid points, and $f(p)$ and $f(\beta)$ are the marginal density functions, computed from the normalized occupation numbers $\tilde w_{ij}$ as in Eq.\ \eqref{eq:marginal}.

First, we computed the CDFs for the marginal DFs, denoted $F_p$ and $F_{\beta}$:
        \begin{equation}
F_{p_m} = \sum_{i=1}^{m} f(p_i) , \quad F_{\beta_n} = \sum_{j=1}^{n} f(\beta_j) .
\label{eq:marginalcdf}
        \end{equation}
        With the CDFs $F_p(S_1)$, $F_p(S_2)$, $F_{\beta}(S_1)$ and $F_{\beta}(S_2)$ computed, we were able to compute the statistical differences as defined in \citet{hn-method}:
        \begin{equation}
\left\{
\begin{aligned}
D_p(S_1, S_2) & = \alpha_k \norm{ F_p(S_1) - F_p(S_2) }_{k} \\
D_{\beta}(S_1, S_2) & = \alpha_k \norm{ F_{\beta}(S_1) - F_{\beta}(S_2) }_{k}
\end{aligned} \quad ,
\right.
\label{eq:popul-diff}
        \end{equation}
        where we computed cases $k=1$, $k=2$ and $k=\infty$, with the scaling factors $\alpha_1 = 1/4$, $\alpha_2 = 1$ and $\alpha_{\infty} = 2$. As a general rule of thumb, the statistical difference between two populations can be considered significant if $D \gtrsim 0.2$. Naturally, one number does not tell everything about the quality of the statistical difference, which is why we used several different norms. Plotting the DFs (and CDFs) of both populations in the same figure is often more illustrative in terms of analyzing differences.

\section{Tests with synthetic data and other examples}

In this section, we demonstrate two applied examples of our method. First, we run a series of synthetic simulations to test the validity of the WISE database. In all of our examples, we used a customized version of the WISE database, which was compiled from the original data in \citet{wise}. The algorithm described in Sect. \ref{sec:synth} was executed several times for different $(p, \beta)$ peak values and population sizes. In the second example, we computed the $(p, \beta)$ distributions for two populations using the algorithm from Sect. \ref{sec:leader-main}, and compared the distributions using the algorithm from Sect. \ref{sec:comparison}. Both examples were executed using MATLAB software \citep{matlab}.

We checked the accuracy of our method by observing how well we were able to reconstruct the peak of the joint $(p, \beta)$ distribution. In our setup, we took a population of $N$ asteroids from DAMIT, with the shape elongation and spin latitude and longitude known for each asteroid model. We used the geometries of the asteroid, the Sun, and the Earth that were computed for the WISE database in \citet{wise}. We observed how accurately the $(p, \beta)$ peak was computed for the population, and repeated this 50 times, each time having a different, randomly generated $(p, \beta)$ peak. This way, we gained a good understanding of how accurately the method computes the peaks for the given population size, no matter where the most occupied bin lies in the $(p, \beta)$-plane. After this, we repeated the same simulation setup for another population size; we considered populations ranging from 100 to 5000 asteroids in order to see how the accuracy of the method improves with a growing number of asteroids in the population.

\begin{figure}
\centering
\includegraphics[width=0.4\textwidth]{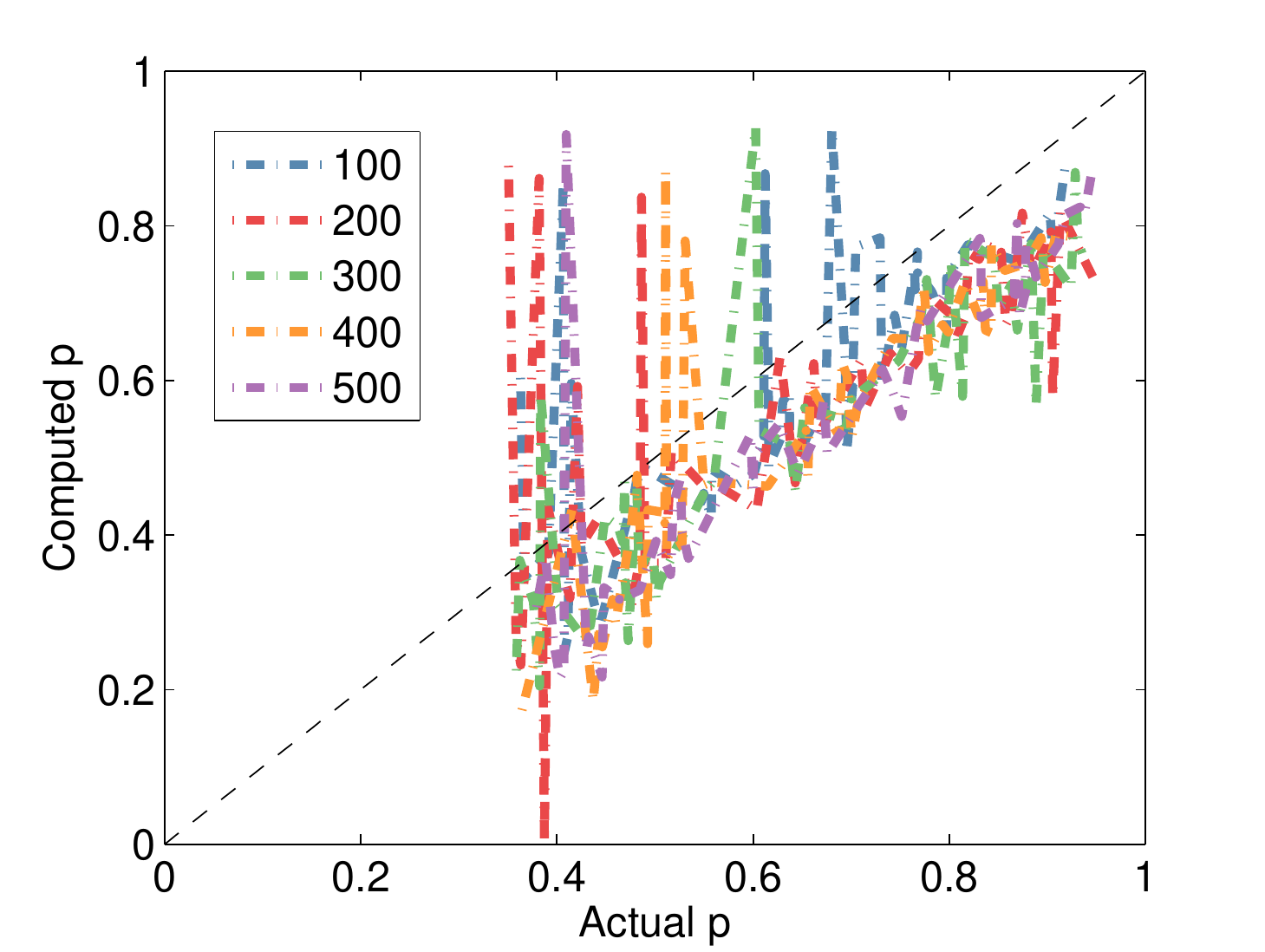}
\includegraphics[width=0.4\textwidth]{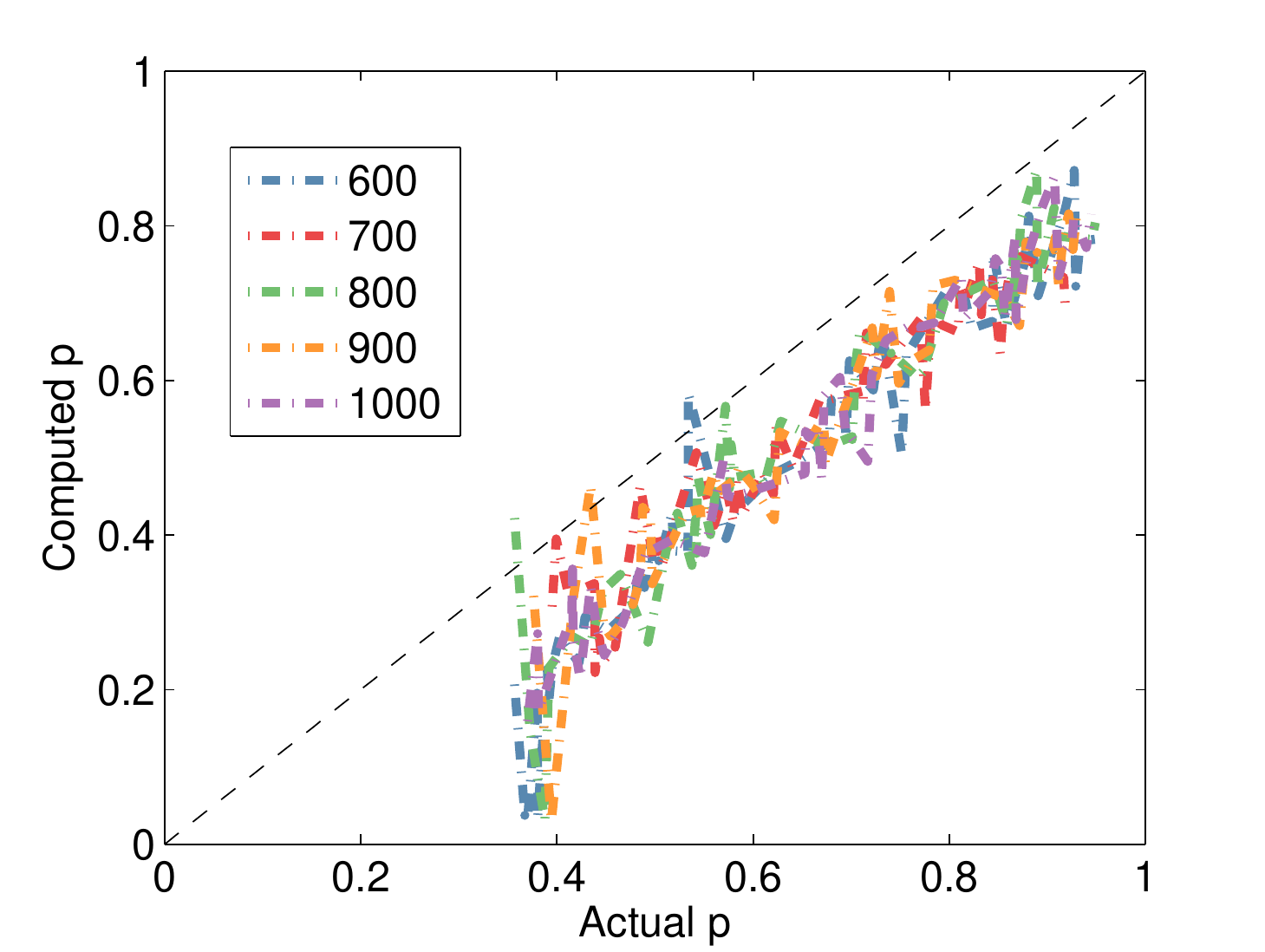}
\includegraphics[width=0.4\textwidth]{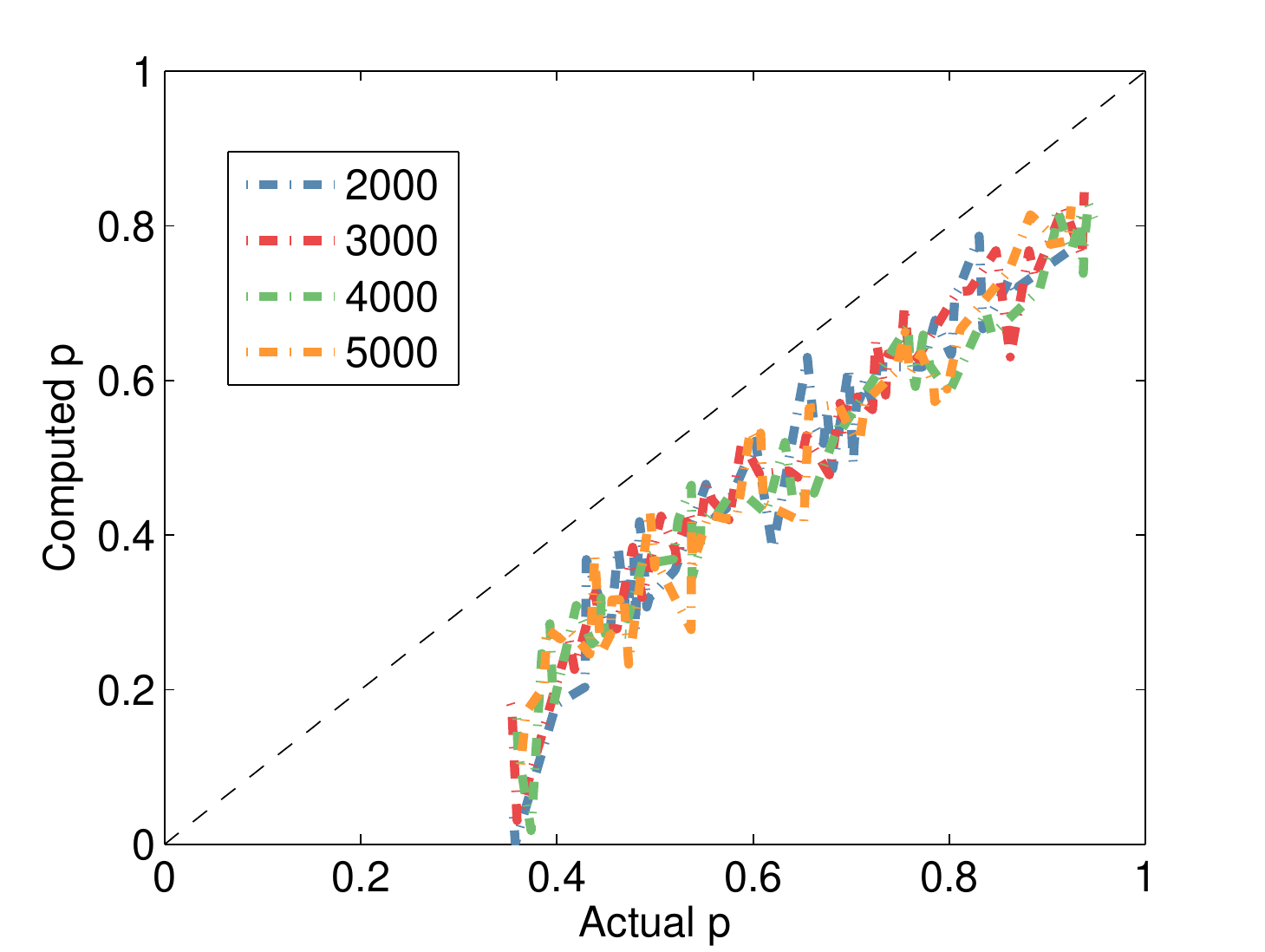}
\caption{Synthetic simulations illustrating how the accuracy of the $p$ solution increases when the population size increases from 100 to 5000 asteroids, with the geometries from WISE. The black, dashed `$y=x$' line presents the ideal case when the computed solution is completely accurate.}
\label{synth-size-p}
\end{figure}

\begin{figure}
\centering
\includegraphics[width=0.4\textwidth]{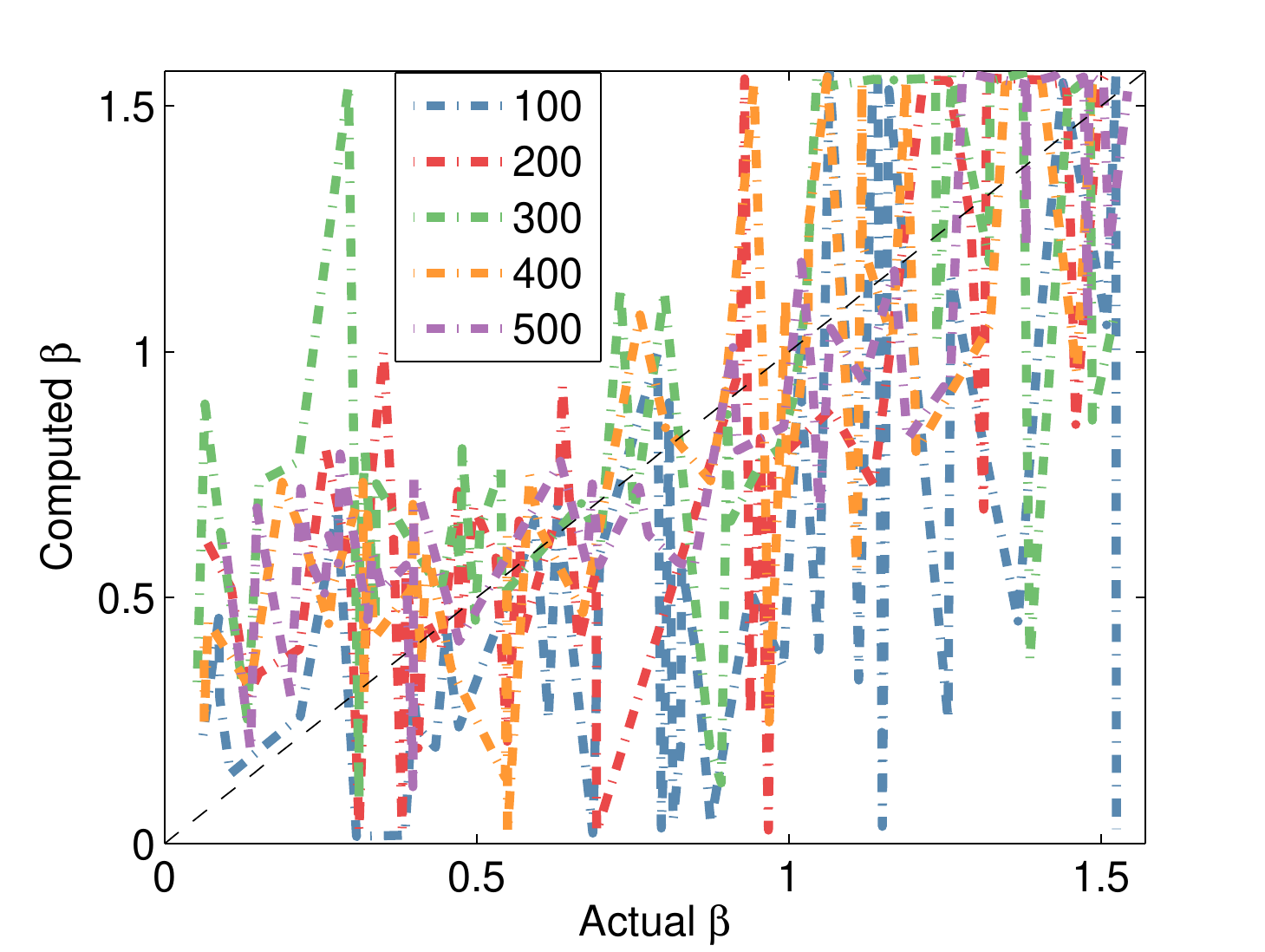}
\includegraphics[width=0.4\textwidth]{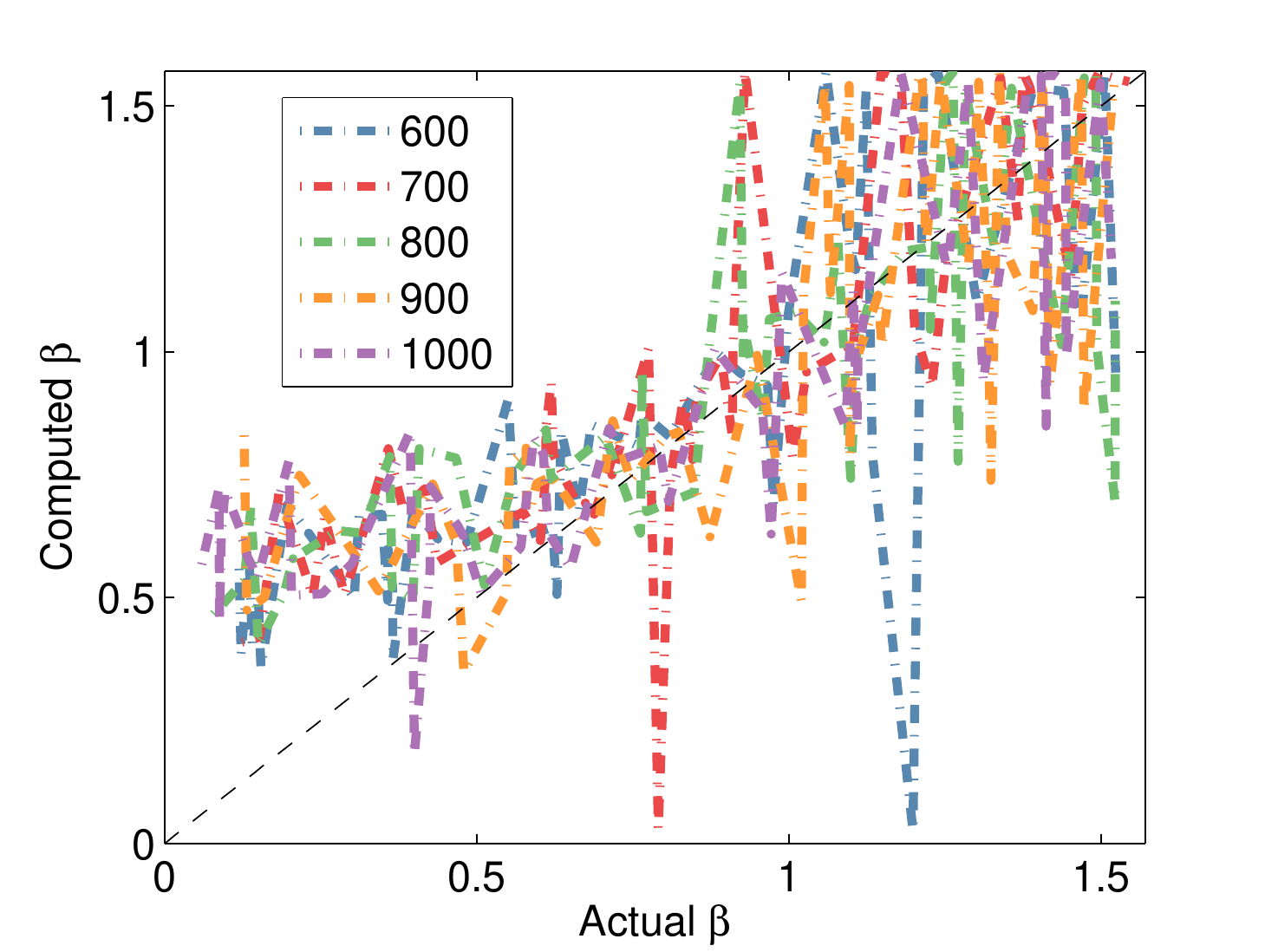}
\includegraphics[width=0.4\textwidth]{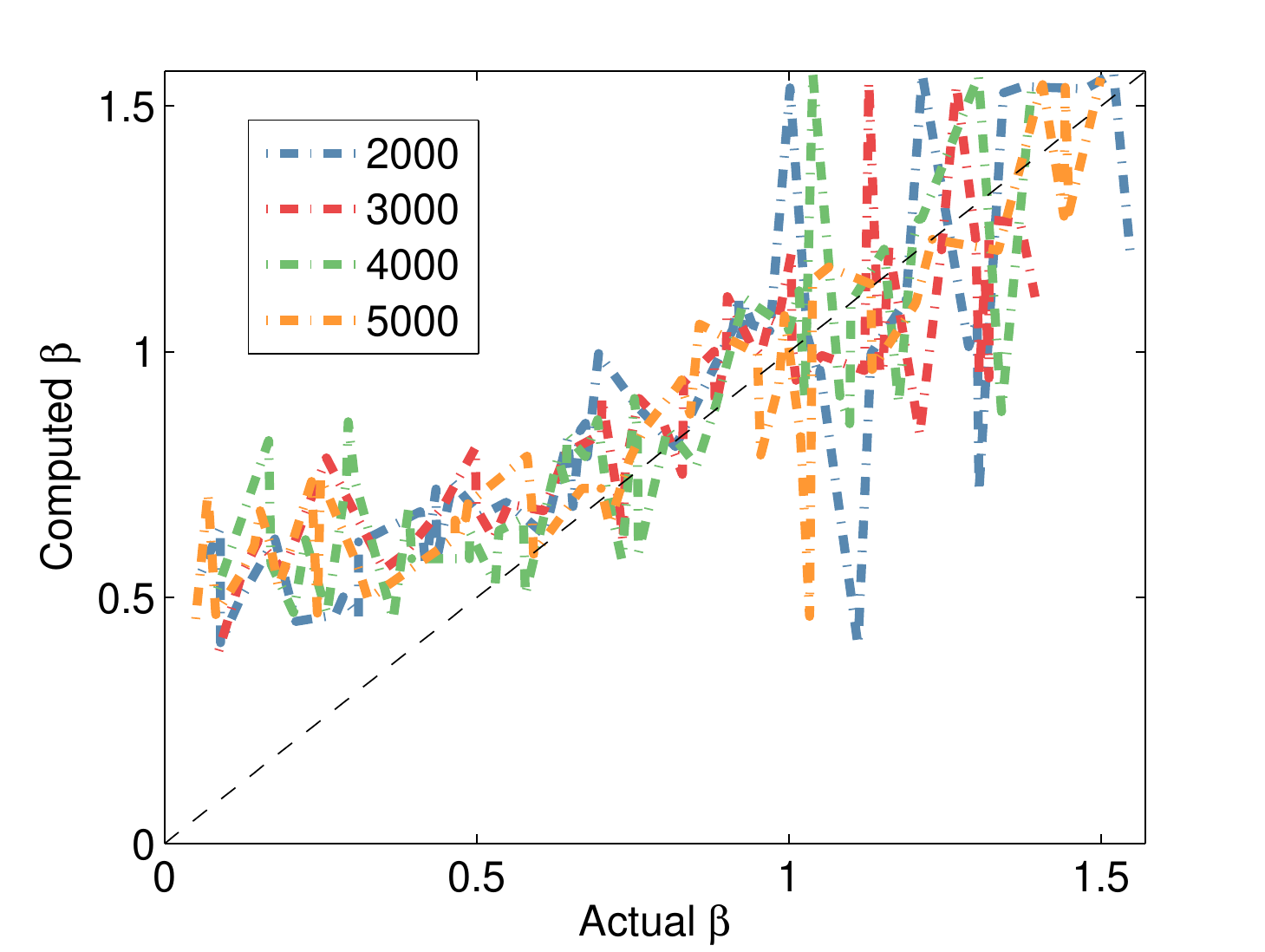}
\caption{Synthetic simulations similar to the ones in Fig.\ \ref{synth-size-p}, but for $\beta$.}
\label{synth-size-beta}
\end{figure}

Plots of different population sizes are presented in Figs. \ref{synth-size-p} and \ref{synth-size-beta}. Each plot draws the actual $p$ or $\beta$ peak versus the computed $p$ or $\beta$ peak, respectively, also showing the ideal case when the solution is completely accurate. As we can see from the plots in Fig.\ \ref{synth-size-p}, the variance in the $p$ peak decreases noticeably when the population size increases. In addition, there is an obvious systematic error that the $p$ peak has been shifted `down', towards more elongated shapes. This shift is mainly due to the model (and data) noise. For example, a spheroidal shape with surface irregularities, estimated at $p=1$, produces photometric variation interpreted as $p<1$ by the smooth ellipsoidal model. For peaks with $p \gtrsim 0.6$, the shift is about 0.1 units, whereas for lower peaks, there is a bigger shift. In realistic populations, shapes with $p < 0.5$ are rare, so it is safe to assume that the value of the $p$ peak is much higher than 0.5. Hence, it usually suffices to expect that the computed $p$ peak is 0.1 units too low for WISE data.

In order to check whether the above result is a bias related to the asteroid database or our way of determining the shape elongation, we repeated the simulations with an alternative way to compute $p$. We considered the contours of the DAMIT-based shapes and computed the ellipsoid (with semiaxes $a$ and $b$, where $a \ge b$) that best fits the contours, and then we computed the shape elongation: $p=b/a$. We observed that this produced no difference in the simulations, so we conclude that the bias is caused by the database; the bias caused by our definition of $p$ is random rather than systematic. From now on, we will use our original definition of $p$ (rather than finding the best elliptical fit for contours), as it is more simple and computationally faster. To correct the systematic error encountered with the WISE database, we shift the $p$ values up by 0.1 units in the deconvolution phase.

As we can see in Fig.\ \ref{synth-size-beta}, the location of the computed $\beta$ peak is a coarse approximation of the actual position. The variance in the accuracy is large. While a bigger population does improve the accuracy of the peak, the variance always remains to some extent. Low $\beta$ values perpendicular to the ecliptic plane have a lower variance, while the high $\beta$ peaks in the ecliptic plane have a high variance, suggesting that the solution is moderately unstable if the actual $\beta$ is high. Since the tail of the computed distribution tends to spread towards the ecliptic plane, as seen in \citet{hn-method}, it is not surprising that the computed peak tends to shift away from the ecliptic plane. Similarly for low $\beta$ values, the computed peak shifts away from the low end of the $\beta$ range. A systematic correction for $\beta$ is complicated to implement, so we merely note that caution should be used with the obtained $\beta$ solution, as our method yields coarse estimates of the spin latitude distribution.

The simulations presented in Figs.\ \ref{synth-size-p} and \ref{synth-size-beta} were done with an equally spaced $(p, \beta)$ grid. We tried simulations where we utilized a lower resolution for different $p$ and $\beta$ values, but we noticed that it reduced the accuracy of the method, especially for $\beta$, when using WISE database. The most accurate results were obtained by using an equally spaced grid. Alternatively, as it is extremely unlikely that $p < 0.4$, we could cut our grid by including only values of $p \in [0.4, 1]$. However, we choose to include the whole interval $[0, 1]$ for completeness, since having high occupation numbers on low $p$ values is usually a good indicator of some systematic error; that is, low $p$ values in the grid are useful for error checking.

For future research, we are interested in testing our method with the Pan-STARRS1 database \citep{ps1}. We tested the accuracy of the method with Pan-STARRS1 using synthetic simulations, and found that the shape elongation $p$ is highly accurate, whereas the solution of the spin latitude $\beta$ is more unstable, as can be expected due to the scatter in the observation geometries. A more detailed analysis of using our method with the Pan-STARRS1 database is presented in \citet{ps1}. We are also interested in whether we can `combine' the databases by taking $\eta$ estimates computed from measurements taken from both WISE and Pan-STARRS1 databases. Therefore, we have performed a preliminary examination on whether our method works and is accurate with combined databases. When we are trying to determine the shape and spin distributions of a certain subpopulation, such as an asteroid family, we may not always have a sufficient number of targets in separate asteroid databases. In such cases, we may attempt to supplement the number of $\eta$s by taking observations from multiple databases, provided our method remains accurate.

\begin{figure}[!ht]
\centering
\includegraphics[width=0.48\textwidth]{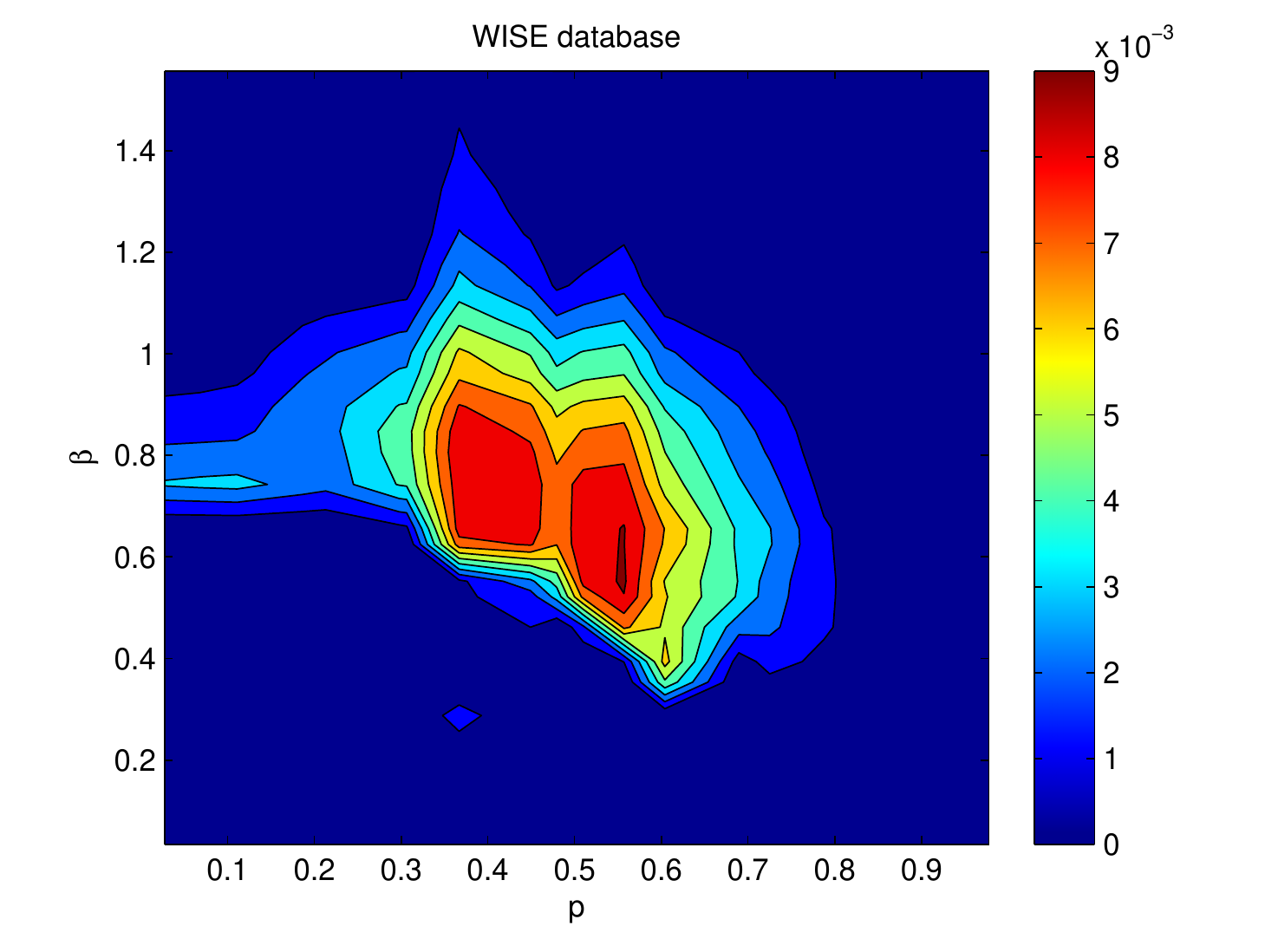}
\includegraphics[width=0.48\textwidth]{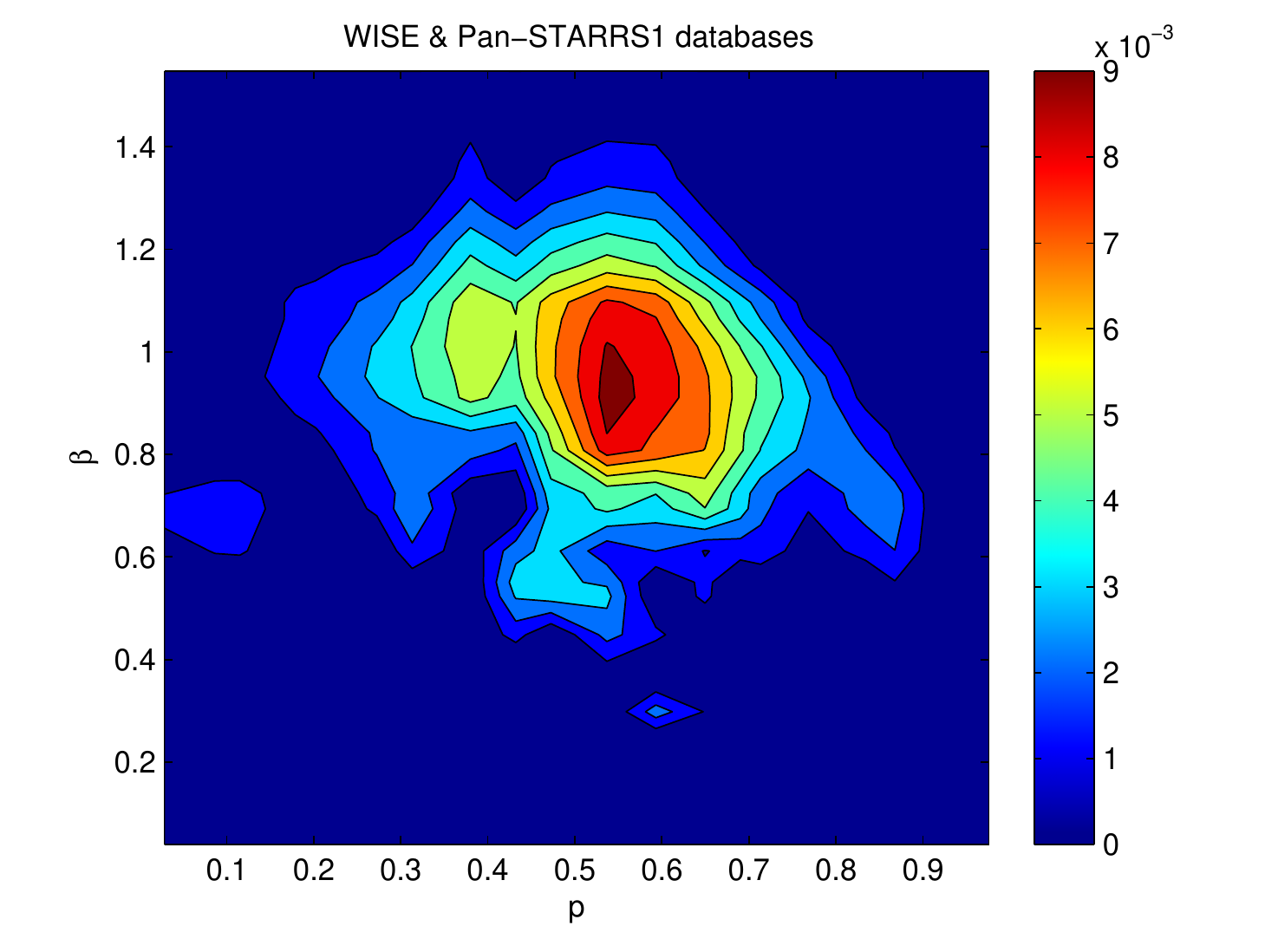}
\caption{Contour solution of the joint $(p, \beta)$ distribution, computed from WISE (top) and combined WISE \& Pan-STARRS1 (bottom) databases.}
\label{fig:WISEvsPS1c}
\end{figure}

\begin{figure}[!ht]
\centering
\includegraphics[width=0.48\textwidth]{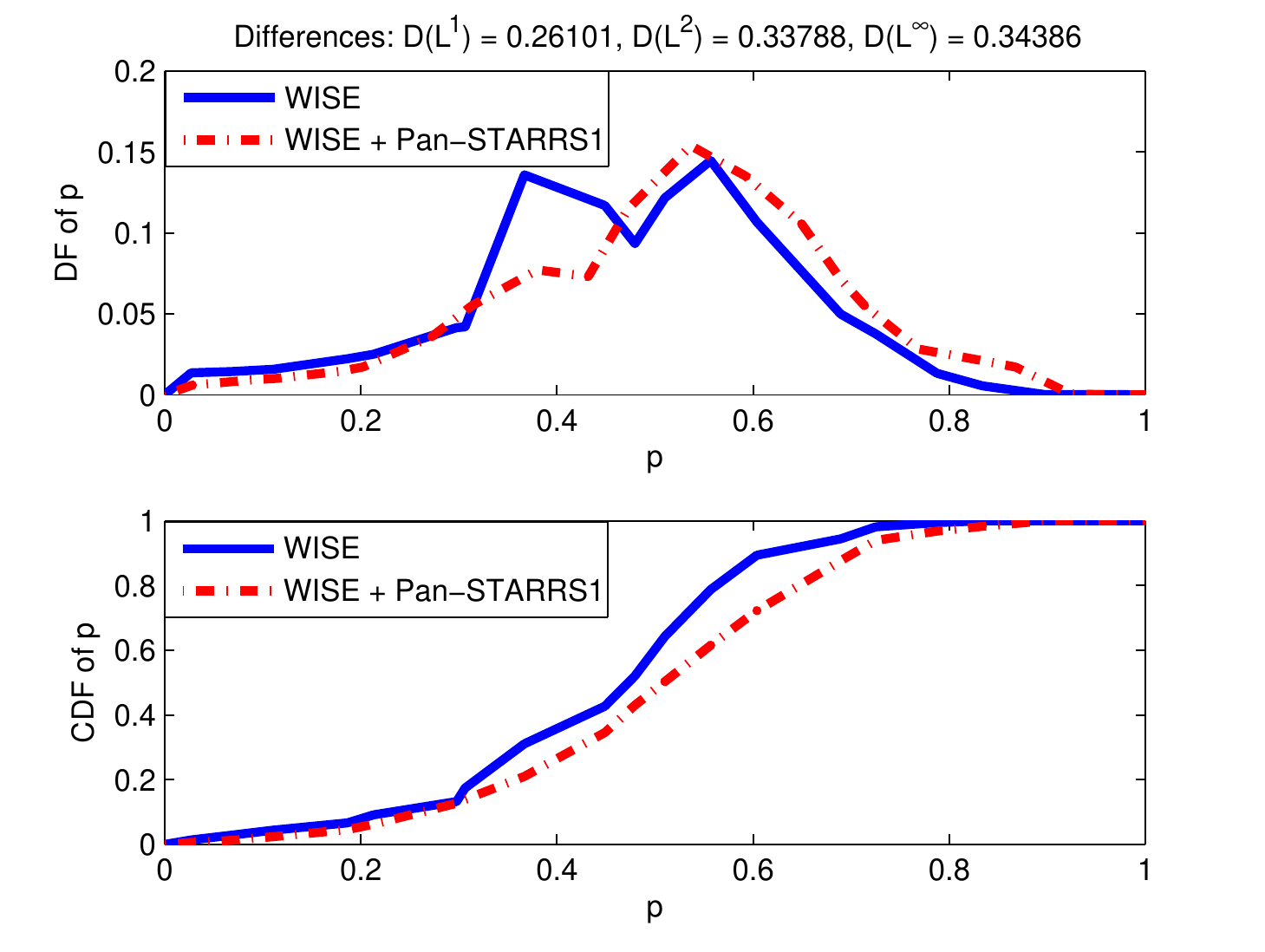}
\caption{Comparison of the marginal DFs (top) for WISE population and combined WISE \& Pan-STARRS1 databases, and of their marginal CDFs (bottom), for shape elongation $p$.}
\label{fig:WISEvsPS1-p}
\end{figure}

\begin{figure}[!ht]
\centering
\includegraphics[width=0.48\textwidth]{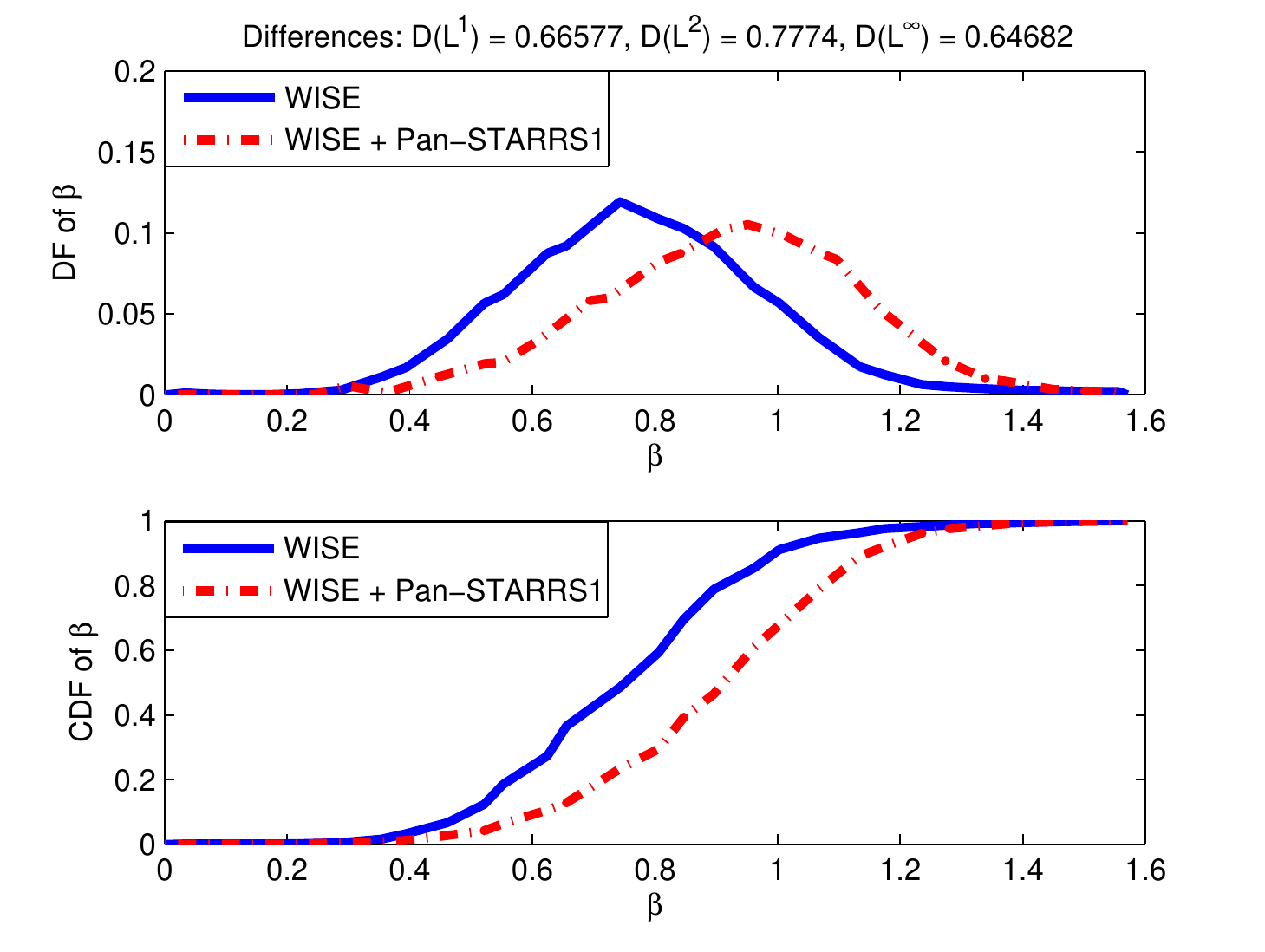}
\caption{Comparison of the marginal DFs (top) for WISE population and combined WISE \& Pan-STARRS1 databases, and of their marginal CDFs (bottom), for spin latitude $\beta$.}
\label{fig:WISEvsPS1-beta}
\end{figure}

To estimate the bias caused by databases, we inverted a subpopulation of about 70,000 asteroids from the WISE database, and compared the results with those obtained from a joint WISE and Pan-STARRS1 sample (containing the aforementioned WISE asteroids, and an additional subset of about 70,000 asteroids from the Pan-STARRS1 database that is used in \citet{ps1}). In the latter sample, we took the brightness variations $\eta$ from each asteroid, and used the combined $\eta$s to construct the CDF $C(A)$. Here we only considered the asteroids that had brightness data available in both databases; this was to avoid selection effects, that is, to ensure that the differences between databases are not caused by observed targets being different. We plotted the computed $(p, \beta)$ distributions in Fig.\ \ref{fig:WISEvsPS1c}, and a comparison of the marginal DFs and their CDFs in Fig.\ \ref{fig:WISEvsPS1-p} for $p$, and in Fig.\ \ref{fig:WISEvsPS1-beta} for $\beta$. No deconvolution has been used for the joint distributions of Fig.\ \ref{fig:WISEvsPS1c}, as we wanted to preserve the information about multiple peaks as well as the spreading behaviour.

Even with half of the population taken from WISE, the combined WISE and Pan-STARRS1 population provides noticeably different distributions compared to merely considering the WISE population. For the marginal distributions of the shape elongation, the differences are minor, despite the moderately high $D(L^i)$ values in Fig.\ \ref{fig:WISEvsPS1-p}. The second double peak is so dominant in the combined database that it dampens the more elongated peak observed from the mere WISE population. In addition, the right tail of the distribution is somewhat heavy when using combined databases. Nevertheless, a visual inspection indicates that these differences in the $p$ distributions are small, and should not be considered significant. Meanwhile, the differences are greater in the marginal distributions of the spin latitude in Fig.\ \ref{fig:WISEvsPS1-beta}, as the $\beta$ peak has been shifted towards the ecliptic plane in the combined population.

\begin{figure}
\centering
\includegraphics[width=0.4\textwidth]{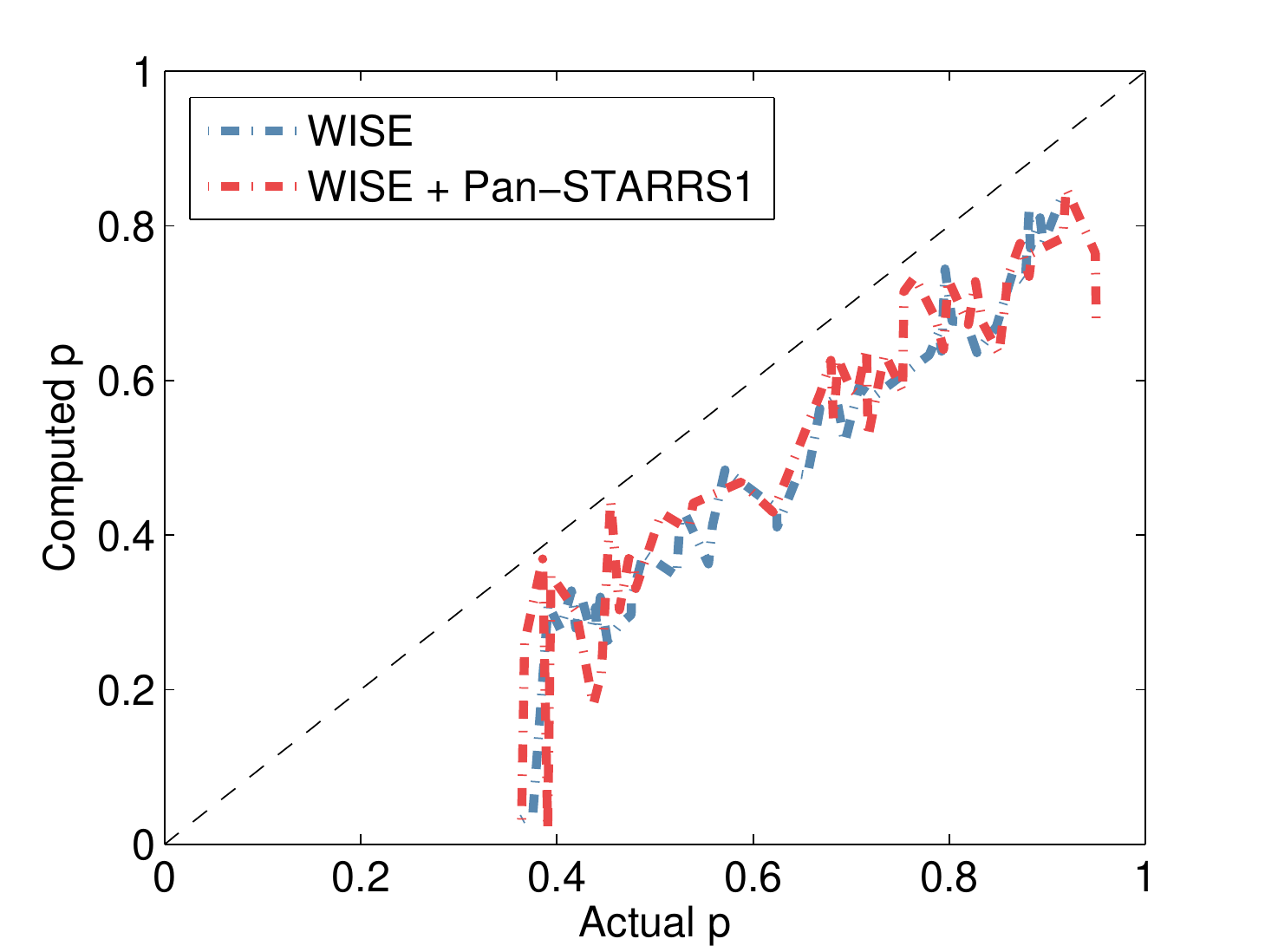}
\includegraphics[width=0.4\textwidth]{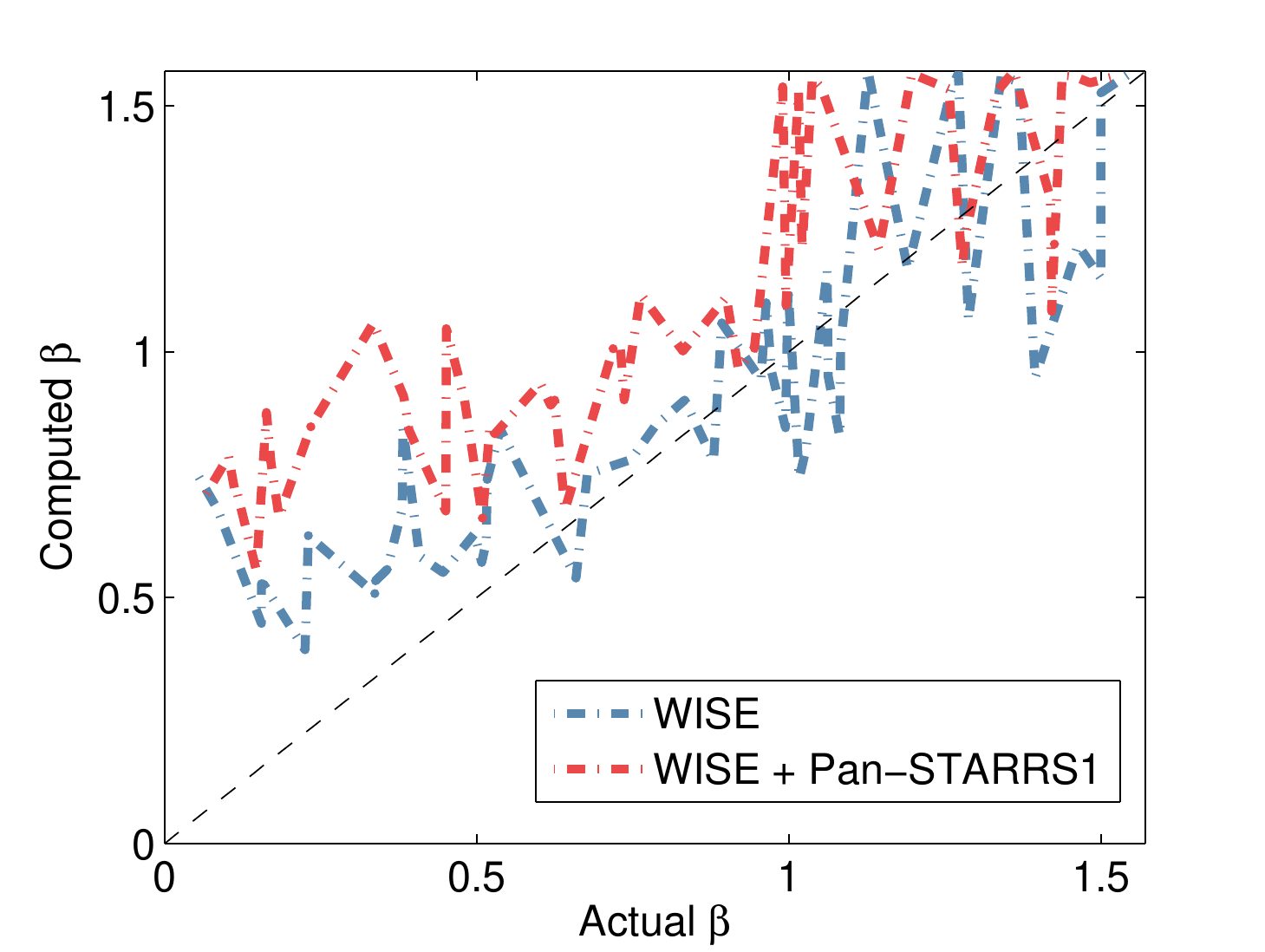}
\caption{Synthetic simulations illustrating the accuracy of the $p$ (top) and $\beta$ (bottom) solutions, using geometries from WISE and a combined WISE \& Pan-STARRS1 database. The black, dashed `$y=x$' lines present the ideal case.}
\label{synth-hybrid}
\end{figure}

To determine whether the differences are caused by our method or the databases, we ran synthetic simulations, using geometries obtained from the above-mentioned databases. Once again, we only took geometries from the targets that had been in observed in both databases. The results from the synthetic simulations have been plotted in Fig.\ \ref{synth-hybrid}, in a format similar to that in Figs. \ref{synth-size-p} and \ref{synth-size-beta}. The accuracy of the $p$ solution is clearly unaffected by the combining of the databases, so the slight differences in the computed $p$ distributions in Fig.\ \ref{fig:WISEvsPS1-p} are likely to be caused by the differences in the databases. As $\beta$ is more sensitive to the observing geometries, the accuracy of the $\beta$ solution deteriorates faster than that of the $p$ solution when we combine databases.

We conclude that combining databases may lead to instabilities, so combining databases may not be useful if they are constructed in a different manner. If a database contains enough targets in a subpopulation, combining it with other databases will not necessarily increase the accuracy of the method. Furthermore, this example shows how much bias the choice of database introduces, which further emphasizes that no database should be blindly used with our method. Instead, every time we are introduced to a new database, it is crucial to run synthetic simulations in order to test the validity and error levels of the method.

\section{Discussion and conclusions}

We have introduced the LEADER software package for the fast estimation of population-level distributions of shape elongations and spin latitudes of asteroids. The method itself is demonstrably robust and designed to extract all the available information from databases that contain at least a few photometric points for each target (and preferably thousands of targets), when the data are not sufficient for individual models of the targets. However, we cannot overemphasize the necessity of testing the usefulness of the database with simulations based on synthetic data at the same geometries as in the database. This is the only way to assess the reliability of the inversion result, which is essentially dictated by the database.

The synthetic simulations provide a good overview of the applicability of the database. If the geometries are undersampled or the number of reasonably good brightness variation estimates is low, the acquisition of useful population-level distribution estimates is not possible regardless of the method. We have also shown how different the results become when switching to another database. For example, when combining the $\eta$ estimates from both WISE and Pan-STARRS1, the bias introduced to the spin latitude is significant compared to using observations from merely the WISE database. As our examples show, some databases are more informative than others even if they all seem to be extensive at a first glance. The simulations are necessary to determine this.

\begin{acknowledgements}

We would like to thank Matti Viikinkoski for his valuable comments and feedback with the software implementation. This research was supported by the Academy of Finland (Centre of Excellence in Inverse Problems). HN was supported by the grant of Jenny and Antti Wihuri Foundation. This publication also makes use of data products from NEOWISE, which is a project of the Jet Propulsion Laboratory/California Institute of Technology, funded by the Planetary Science Division of the National Aeronautics and Space Administration. In addition, this research made use of the NASA/IPAC Infrared Science Archive, which is operated by the Jet Propulsion Laboratory/California Institute of Technology, under contract with the National Aeronautics and Space Administration. We would also like to thank the reviewer for insightful comments that improved this paper. Last but not the least, we would like to thank Josef \v{D}urech and his research group for providing important feedback concerning the software package.

\end{acknowledgements}


\begin{appendix} 

\section{Structure of the software package}

In this appendix, we present the overall structure of LEADER as a rough-scale pseudocode, as well as listing the key functions of the software package. First we discuss the main routine, then the variant used for synthetic simulations, and the application for comparing distributions. The software package presented here is designed for analyzing the WISE database.

The database is available for download in DAMIT, under the Software section\footnote{http://astro.troja.mff.cuni.cz/projects/asteroids3D/web.php?page=download\_software}.

\subsection{Main routine}

\begin{table}%
\begin{Verbatim}[frame=single]
function leader_main_WISE

for i=1:N
   call function lcg_read_WISE(datafile i)
      read datafile
      split data into measurement sets
      compute eta for each set
      compute A from eta
      return A
   endfunction
end

A_sort = sort(A_vec)
for j=1:length(A_sort)
   CDFA(j) = j/length(A_sort)
end

call function leader_invert
   create (p, beta) grid
   create Mtilde, Ctilde
   W = lsqnonneg(Mtilde, Ctilde)
   return W
endfunction

call function leader_plots
   plot contour in (p, beta) plane
   compute marginal distributions
   draw marginal distributions for p and beta
endfunction

call function leader_postprocess_WISE
   find (p, beta) peak
   dampen bins away from the peak
   P_new = P + deltaP
   draw deconvoluted contour in (p, beta) plane
endfunction

endfunction
\end{Verbatim}
\caption{Pseudocode of the main function of LEADER.}
\label{tab:main}
\end{table}

\noindent The principle of the main function \verb+leader_main_WISE+ has been presented on Table \ref{tab:main}. We assumed the population consists of $N$ asteroids, and for each asteroid, we have a datafile available. For each datafile, we read the geometries, measurement times and brightness intensities using the function \verb+lcg_read_WISE+. We split the observations into multiple sets according to the principles mentioned in Sect. \ref{sec:forward}: all measurements in the same set must be done within a small enough change in geometry, and at least five observations are required. Then for each set, we computed the brightness variation $\eta$ and amplitude $A$ using Eq.\ \eqref{eq:eta-A}. After analyzing all datafiles, we had a list of amplitudes (the vector \verb+A_vec+ on Table \ref{tab:main}). We sorted the elements of the vector into an increasing order, and then we construct the CDF of $A$. In the inversion phase (function \verb+leader_invert+), the matrix $\tilde M$ and the vector $\tilde C$ are generated as explained in Sect. \ref{sec:inverse}. The Matlab function \verb+lsqnonneg+ computes the solution to the non-negative least-squares problem
$$ \min_w \norm{\tilde M w - \tilde C}_2^2 , \quad \text{where } w \ge 0 . $$
Finally, we used the function \verb+leader_plots+ to plot the solutions, and the function \verb+leader_postprocess_WISE+ to deconvolute the solution visually.

\subsection{Synthetic simulator}

\begin{table}%
\begin{Verbatim}[frame=single]
function leader_synth_main_WISE

for i=1:N
   set p_wanted, beta_wanted, lambda_wanted
        
   while ( |p-p_wanted| > tol )
      call function damit_model
         read random datafile
         return vertex and face information
      endfunction
      apply stretch on vertices
      call function leader_ellipsoid
         compute a, b, c, p = b/a
         return p
      endfunction
   endwhile

   call function
   leader_brightness_synth_WISE(datafile i)
      beta=beta_wanted, lambda=lambda_wanted
      call function lcg_read_synth_WISE
         read datafile
         return dates, geometries
      endfunction
      compute L for each geometry, add noise
      split data into measurement sets
      compute eta for each set
      compute A from eta
      return A
   endfunction
end

A_sort = sort(A_vec)
for j=1:length(A_sort)
   CDFA(j) = j/length(A_sort)
end
plot synthetic contour in (p, beta) plane

call function leader_invert
   create (p, beta) grid
   create Mtilde, Ctilde
   W = lsqnonneg(Mtilde, Ctilde)
   return W
endfunction

call function leader_plots
   plot contour in (p, beta) plane
   compute marginal distributions
   draw marginal distributions for p and beta
endfunction

call function leader_postprocess_WISE
   find (p, beta) peak
   dampen bins away from the peak
   P_new = P + deltaP
   draw deconvoluted contour in (p, beta) plane
endfunction

endfunction
\end{Verbatim}
\caption{Pseudocode of the synthetic simulator implementation of LEADER.}
\label{tab:synth}
\end{table}

\noindent We have presented the principle of the synthetic simulator \verb+leader_synth_main_WISE+ on Table \ref{tab:synth}. We assumed a population of $N$ asteroids. For each target, we fixed a desired $(p_{\text{wanted}}, \beta_{\text{wanted}})$ value, and we chose the longitude $\lambda_{\text{wanted}}$ from a random uniform distribution $[0, 2\pi]$. Then, we kept selecting randomized and stretched asteroid models from the DAMIT database, until we had an asteroid with the desired $p$ value (within a certain tolerance). Then we read measurement dates and geometries (direction of the Sun and Earth) from the database we were testing (for example, WISE). We fixed $\beta = \beta_{\text{wanted}}$ and $\lambda = \lambda_{\text{wanted}}$, and used them to transform the direction vectors of the Sun and Earth into the asteroid's own frame. Then, for each geometry, we used a scattering law to compute the brightness intensity $L$, adding a small Gaussian noise. The rest of the algorithm is identical to the main routine: we split the data into measurement sets, computed $\eta$ and $A$ for each of them, constructed the CDF $C(A)$, used the subfunction \verb+leader_invert+ to compute the solution distribution, and finally visualized the solution with \verb+leader_plots+ and \verb+leader_postprocess_WISE+ subfunctions. It is recommended that some fine-tuning is done in the deconvolution function (\verb+leader_postprocess_WISE+) to reshape the solution closer to the synthetic $(p, \beta)$ distribution from the forward model.

\subsection{Comparison of asteroid populations}

\begin{table}%
\begin{Verbatim}[frame=single]
function ast_comparison_WISE

load population1
call function leader_main_WISE
   return p1, beta1, fp1, fbeta1
endfunction

load population2
call function leader_main_WISE
   return p2, beta2, fp2, fbeta2
endfunction

call function KS_comparison
   create CDFs of fp1, fp2, fbeta1, fbeta2
   % CDFs are called Cp1, Cp2, Cb1, Cb2
   Cp2i = interpolate Cp2 at p1
   Cb2i = interpolate Cb2 at beta1
   for k=[1, 2, inf]
      Dp = alpha_k*||Cp1-Cp2i||_k
      Db = alpha_k*||Cb1-Cb2i||_k
   end
   plot margin distributions in the same figure
   plot CDFs in the same figure
   return Dp, Db
endfunction

endfunction
\end{Verbatim}
\caption{Pseudocode of the statistical comparison of two asteroid populations.}
\label{tab:comparison}
\end{table}

\noindent The application for the statistical comparison of asteroid populations is called \verb+ast_comparison_WISE+, and its principle has been presented on Table \ref{tab:comparison}. We ran the main routine \verb+leader_main_WISE+ first for population 1, saving the used $p$ and $\beta$ grids and their marginal DFs into variables \verb+p1+, \verb+beta1+, \verb+fp1+ and \verb+fbeta1+. We doidthe same for population 2, saving the grids and their marginal DFs respectively into variables \verb+p2+, \verb+beta2+, \verb+fp2+ and \verb+fbeta2+. Then we called the subfunction \verb+KS_comparison+. The subfunction constructs the CDFs of each marginal DF, with \verb+Cp1+ and \verb+Cp2+ being the CDF of $p$ for populations 1 and 2, and \verb+Cb1+ and \verb+Cb2+ being the CDF of $\beta$ for populations 1 and 2, respectively. To compare the CDFs, we interpolated \verb+Cp2+ and \verb+Cb2+ at the grid points of population 1. Then we computed the statistical differences as defined by Eq.\ \eqref{eq:popul-diff}. Finally, we plotted the marginal DFs and their CDFs as in Figs. \ref{fig:WISEvsPS1-p} and \ref{fig:WISEvsPS1-beta} to illustrate the differences of the distributions.

\end{appendix}

\end{document}